# Sequential Operational Decision-Making for Power System Resilience Under Evolving Wildfires

Arastoo H Salimi, Majid Dehghani, *Student Members, IEEE*, Hamidreza Nazaripouya, *Senior Member, IEEE*

***Abstract*— This paper proposes a novel automated decision-support framework aimed at enhancing the resilience of power systems and operational resilience against wildfires by formulating the decision-making process as a stochastic multi-stage programming during a progressive wildfire. The paper develops a framework that takes into account both preventive and corrective actions, enabling automated and adaptive decisions based on potential scenarios over the course of a wildfire's progression. This approach considers the evolving nature of the wildfire threat and seeks to optimize the response strategies accordingly throughout its duration. The objective is to minimize wildfire risk and operational costs while reducing load curtailment. The framework accounts for potential contingencies caused by progressive wildfires. A novel algorithm is proposed to construct a decision tree based on wildfire progression and system geographical information. Additionally, a novel stochastic dual dynamic programming approach is deployed to solve the proposed optimization problem, achieving a global optimum for the framework. The effectiveness of the proposed method is demonstrated on the IEEE 30-bus system under various wildfire impact scenarios and is then applied to the IEEE 300-bus system to illustrate the scalability of the proposed approach. The results highlight the advantages of the proposed automated framework over a single-stage operational optimization strategy in enhancing power grid resilience under wildfire conditions.**

***Indices and Sets***

| | |
|---|---|
| $e \in E$ | Set of lines. |
| $B_i^e$ | Denotes the subset of lines that connected to bus $i$. |
| $g \in G$ | Set of generators. |
| $i \in B$, $j \in B$ | Set of buses. |
| *a(n)* | Denotes the parent node of $n$. |
| *c(n)* | Denotes the set of child nodes of $n$. |
| *D, G, L,* and *B* | Represent the sets of load demand, generators, transmission lines, and buses, respectively. |
| $B_i^D$, $B_i^G$, $B_i^e$ | Represent the set of loads, generators, and lines connected to bus *i*, respectively. |
| $\omega_{d,n}$ | Weight to express that certain load, or service may be prioritized over others, in node *n*. |
| $\alpha$ | $\alpha \in [0 \;\; 1]$ represents the trade-off between prioritizing load service and cost reduction (low $\alpha$) and avoiding wildfire risk (high $\alpha$). |

***Parameters and constants***

| | |
|---|---|
| $b_{e,n}$ | Susceptance of branch e (line) in the network at node *n*. |
| $D_{i,n}^P$ | Real power demand at bus *i* (MW), at node *n*. |
| $P_g^{max}$, $P_g^{min}$ | Maximum and minimum real power generation limit at generator $g$ (MW). |
| $R_{fire,n}$ | Denotes the risk of wildfire in node *n*. |
| $D_{Tot,n}$ | Represents the total amount of load delivered in node *n*. |
| $R_{d,n}$ | The wildfire risk associated with load *d* at node *n* in a specific area. |
| $R_{g,n}$ | The wildfire risk associated with generator *g* at node *n* in a specific area. |
| $R_{e,n}$ | The wildfire risk associated with line *e* at node *n* in a specific area. |
| $R_{i,n}$ | The wildfire risk associated with bus *i* at node *n* in a specific area. |
| $C_{g,n}$ | Function to calculate the associated cost for each generator at node *n*. |
| $\theta^{max}$, $\theta^{min}$ | Additional constants added to the angle differences. |
| $x_*$, $y_*$ | Refer to the geographical location of systems, including power system lines and components. |

***Notation for Optimization Problem***

| | |
|---|---|
| $p_{g,n}$ | Real power dispatched at generator *g* (MW) at node *n*. |
| $P_{e,n}$ | Real power flow on branch *e*, at node *n*. |
| $\theta_{i,n}$ | Voltage phase angle at bus $i$ (radians), at node *n*. |
| $\rho_{nm}$ | The transition probability that node *m* is realized conditional on its parent node *n* being realized. |
| $l_{d,n}$ | $l_{d,n} \in [0,1]$ that represents the fraction of the load (and corresponding distribution infrastructure) that is de-energized in node *n*. |
| $u_{g,n}$ | $u_{g,n} \in \{0,1\}$ indicate whether a generator *g* is energized or not in node *n*. |
| $u_{e,n}$ | $u_{e,n} \in \{0,1\}$ indicate whether a line *e* is energized or not in node *n*. |
| $u_{i,n}$ | $u_{i,n} \in \{0,1\}$ indicate whether a bus *i* is energized or not in node *n*. |
| $v_n$ | Vector of decision variable in node *n*. |
| $w_n$ | Vector of internal variable in node *n*. |
| $f_n$ | The nonnegative nodal cost function of the problem at node *n*. |
| $Q_n(v_{a(n)})$ | The value function of node *n*. |
| $F_n$ | Feasibility set in some Euclidean space of decision variables $(v_n, w_n)$ of the nodal problem at node $n$. |
| $V_n$ | Denote the image of the projection of $F_n$ onto the subspace of the variable $v_n$. |
| $z_n$ | Local variable of node $n$ and impose the duplicating constraint $v_{a(n)}=z_n$. |
| $\psi_n$ | Lipschitz continuous penalty function. |
| $\sigma_n$ | Coupling constraint. |
| $\lambda, \rho$ | Compact set of parameters. |
| $C_m$ | Generalized conjugacy cut for $Q_m$. |

# I. Introduction

IN recent years, the severity and frequency of natural disasters, particularly wildfires, have escalated due to global warming. Rising temperatures have extended and intensified wildfire seasons in the United State regions, where prolonged drought conditions have significantly increased the risk of fires. Research indicates that anthropogenic climate change has already doubled the area of forest burned in recent decades [1].

Electrical infrastructure is among the most vulnerable systems impacted by wildfires. In rural regions, wildfires pose significant threats to sections of the transmission grid, leading to widespread power outages that severely impact the well-being of electricity consumers. Furthermore, these disruptions result in substantial financial losses. In Texas, two wildfires ignited in Bastrop County in 2011 due to trees making contact with power lines. These fires became the most destructive in Texas's history, resulting in four fatalities and causing over $300 million in damage [2]. In 2019, an unprecedented peak expenditure was recorded, with the U.S. Forest Service and other federal agencies allocating over $2.9 billion to address wildfire challenges [3]. Wildfires create significant challenges for grid operations, as they not only threaten power infrastructure but can also be triggered by electrical faults from power equipment [4]. Since 2015, power lines have been responsible for six of California's 20 most destructive wildfires [5]. From 2014 to 2022, equipment belonging to California's three major utility companies caused over 4,946 fires. Notably, wildfires initiated by power lines and electrical equipment are among the most destructive ones. California's second-largest wildfire ignited when power lines came into contact with a tree. The fire, which began on July 13, 2021, burned 963,309 acres across five counties in Northern California before being contained [6].

Significant efforts have been made in recent years to enhance grid resilience through infrastructure hardening. However, these passive efforts are extremely costly. Fortunately, these costs can be minimized or delayed by incorporating automation and smart control measures into the operational procedures of the system (i.e., operational resilience) to effectively manage resources/assets, and survive a crisis as it happens.

Broadly, the operational resilience of power systems against wildfires can be classified into preventive, corrective, and restorative measures.

A portion of the research in this domain concentrates on operational preventive measures. The study [7] has proposes an optimization model to assist in short-term preventive decision-making under conditions of high wildfire risk. This model is aimed at optimizing grid operations to maximize power delivery while proactively reducing the risk of wildfires by selectively de-energizing grid components. In this study, the optimization model proactively considers both the risk of wildfires and the impact of power outages when optimizing power system operations. Authors in [8] model wildfire events as stochastic disruptions characterized by random magnitude and timing. Their stochastic program aims to maximize electricity delivery while proactively de-energizing components over multiple time periods to mitigate wildfire risks. They employ a cellular automaton model to simulate the initiation and propagation of exogenous and endogenous wildfires based on environmental data. Reference [9] introduces an optimization problem focused on reconfiguring networked microgrids to both manage wildfire risk and maximize the power supplied to customers. This approach also incorporates equity by evaluating customers' ability to cope with power outages, as measured by their social vulnerability. Authors in [10] propose an alternative approach to Public Safety Power Shutoffs (PSPS) by incorporating endogenous uncertainty and utilizing an optimization model. This model determines appropriate network topology changes through switching actions to reduce power flow through vulnerable grid sections, thereby decreasing the probability of wildfire ignition.

Several studies emphasize corrective measures for power systems in response to wildfires. The authors in [11] propose a stochastic programming approach to enhance the resilience of a distribution system exposed to wildfires, considering dynamic line rating of overhead lines. It employs a mixed-integer quadratic optimization formulation to optimally manage and coordinate all local energy resources, aiming to reduce load outages and enhance system resilience during wildfires. Reference [12] proposes a stochastic optimal strategy for equitable corrective actions in distribution grids. This approach aims to enhance the operational resilience of power systems against progressive wildfires while considering equitable distribution of power outages in unbalanced networks. The authors in [13] develop a mixed-integer conic optimization framework that models the progressive thermal impacts of wildfires on overhead conductors and coordinates network reconfiguration and backup generation to minimize load curtailment while maintaining line thermal limits. The authors in [14] propose a nonlinear multi-objective optimization approach to determine the optimal operational strategy for enhancing grid resilience as the primary objective, while considering equity as the secondary objective. Reference [15] propose a probabilistic proactive generation redispatch strategy, utilizing a Markov decision process, to improve the operational resilience of power grids during wildfires.

Besides the limited research conducted on operational grid resilience solutions against wildfires, existing state-of-the-art models in the literature tend to oversimplify the problem. They often ignore the progressive behavior of wildfires, system limitations such as the ramping constraints of power generation units, and the dependency between corrective strategies and preventive decisions (e.g., PSPS). For more accurate analysis, preventive action decisions should be integrated into the optimization process throughout the entire wildfire period. In wildfire management, the sequential nature of decision-making causes operational choices at each step to influence the actions that follow.

As a result, preventive and corrective decisions must be integrated into a unified automated multistage programming framework. In this approach, decision variables such as power generation and preventive measures are made "here and now" choices, while corrective decisions are influenced by prior choices and are made "wait and see" under the uncertainty of wildfire spread scenarios. This framework automates sequential decisions under uncertainty and reduces reliance on manual operator interventions.

Constructing an accurate decision tree to capture the evolution of future system scenarios is a fundamental requirement in multistage optimization. In addition, solving a multistage stochastic program remains highly challenging, especially when mixed integer variables are involved, as the computational complexity increases exponentially with the planning horizon and system size. Various approaches are employed to solve stochastic multistage optimization problems, including stochastic programming [16]-[17], robust optimization [18]-[19], online optimization [20], dynamic programming [21], or combinations thereof [22].

A dynamic programming approach with novel algorithms is employed in this study to tackle multistage stochastic mixed-integer nonlinear programs (MSMINLP). These algorithms, rooted in stochastic dual dynamic programming (SDDP), integrate nested decomposition, deterministic and stochastic sampling, and cuts derived from generalized conjugacy. They enable the identification of global optima without necessitating complete recourse, thus marking a substantial advancement over traditional SDDP methodologies [23].

The contribution of this paper can be summarized as

1- Developing mathematical multistage optimization models that more accurately represent real-world decision-making processes for automated operational grid resilience against wildfires through integrating both preventive and corrective actions into the optimization process. The model acknowledges that decisions made for preventive measures influence decisions in future stages. The proposed approach incorporates these sequential decisions by treating PSPS actions as state variables, considering their impact on future states, thereby ensuring effective decision-making throughout the optimization.
2- Proposing a novel algorithm for wildfire-driven decision tree construction under wildfire spread scenarios, enabling adaptive and automated decision-making based on the system's geographical information.
3- Employing a novel decomposition algorithm for stochastic dual dynamic programming (SDDP) to solve the multistage optimization problem associated with wildfire management with large number of mixed-integer variables. By utilizing this algorithm, it is ensured that the global optimum is accurately determined for the proposed problem.

The rest of the paper is organized as follows:

In Section II, the preventive-corrective multistage optimization problem for grid resilience against wildfires is formulated. Section III presents the wildfire-driven decision tree construction algorithm. In Section IV, the SDDP algorithm is derived, and the solution approach is outlined, detailing the principles of the proposed method. Section V presents the optimization results for the IEEE 30-bus system, along with a scalability demonstration performed on the IEEE 300-bus system. Finally, Section VI provides concluding remarks.

## II. Problem Formulation

This section presents a comprehensive formulation of the multistage optimization problem concerning automated operational grid resilience against wildfire, with the primary objectives of minimizing operational costs, load curtailment, and wildfire risk

The propagation properties and spatiotemporal characteristics of wildfires have unique impacts on the performance of power system components. As wildfires spread, different components may be affected at successive time intervals. Wildfires are inherently stochastic due to the uncertainty in their spread, intensity, and the likelihood of being extinguished at any moment. Wildfire trajectories and their impact on power grid components are unpredictable, as different potential paths can influence system operations differently. The limited ability of power systems to adapt to dynamic wildfire conditions complicates decision-making.

Relying solely on the current states of the system, without accounting for the spectrum of potential future scenarios, results in decisions that are suboptimal under progressive wildfire conditions. To address this, operational resilience must be framed as a multistage stochastic optimization problem, where sequential decisions are tightly interdependent and early choices impose lasting constraints on subsequent actions. One of the foremost difficulties in this formulation is the construction of a decision tree that can faithfully capture the branching progression of a wildfire under different spread scenarios and accurately reflect real operating conditions. Another critical aspect is the definition of state and internal variables, which must be carefully specified to ensure an accurate and coherent representation of the system's evolving conditions.

Wildfires affecting power systems can be classified into two types: exogenous and endogenous [8], Exogenous wildfires are caused by external factors beyond the control of grid operators, while endogenous wildfires result from faults within the system's components. These two types of fires cannot be analyzed in isolation when optimizing power flow, particularly in regions with a high risk of wildfires. Even if the risk of endogenous fires is mitigated, exogenous wildfires may still occur. Therefore, a comprehensive framework must jointly incorporate preventive and corrective actions, recognizing their interdependence and ensuring that operational strategies are not only robust against immediate contingencies but also adaptive to the unfolding and uncertain progression of wildfire events.

For endogenous fires, system operators can reduce risk by assessing potential hazards and selectively de-energizing components. In optimizing preventive actions, the primary objectives are to minimize wildfire risk ($R_{fire}$). However, preventive decisions influence the future operational state of the power system during a wildfire event, potentially altering power flow patterns. Thus, a comprehensive approach is needed to account for the interdependence of decisions over time. Corrective actions during wildfires typically aims to minimize load curtailment $D_{Tot}$ and reduce the operational costs of power generation $C_g(p_g)$. This enables the system to adapt and recover once wildfires, whether of endogenous or exogenous origin, disrupt network operation.

A multistage preventive-corrective optimization framework is proposed to effectively capture the progressive nature of wildfire events. This introduces the index $n$ in the formulation, which refers to the nodes of the multistage optimization decision tree used in the optimization process. To avoid confusion, whenever we refer to "nodes" in this paper, it specifically denotes the decision tree nodes, not the busbars of the power system. When discussing busbars, we will explicitly refer to them by their bus numbers to maintain clarity.

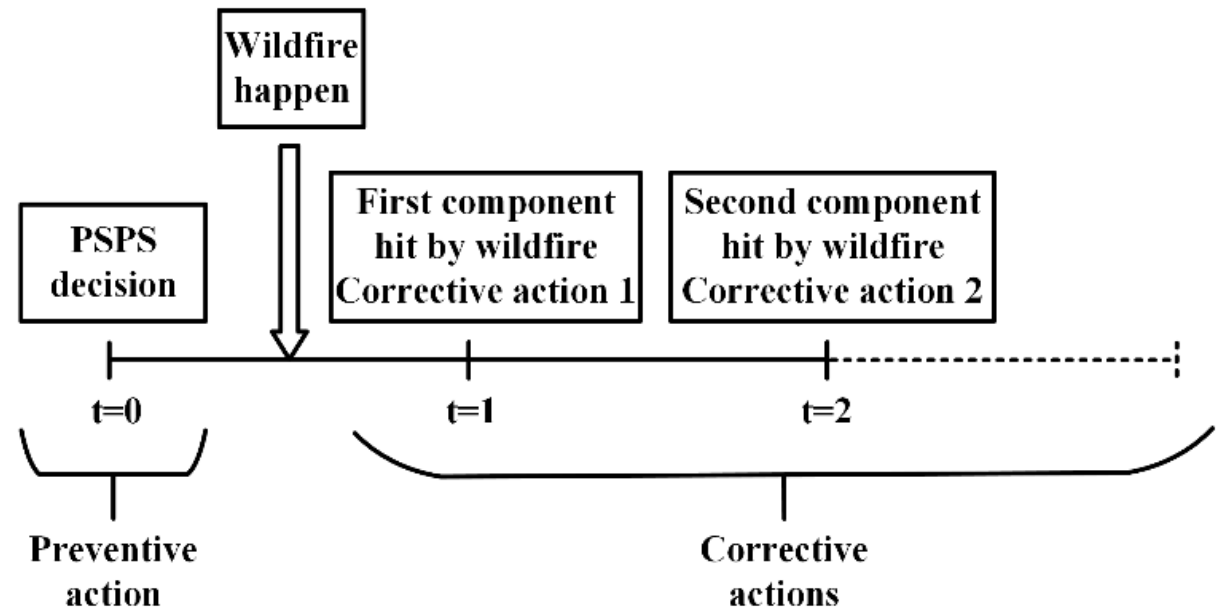

**Figure 1** Multistage optimization framework integrating preventive and corrective decisions for wildfire resilience.

As illustrated in Figure 1, within the multistage optimization framework, preventive decisions are determined in advance of potential endogenous or exogenous wildfire occurrences in subsequent stages. For making PSPS decisions, the formulation incorporates not only the immediate wildfire risk but also considered a spectrum of possible future fire spread scenarios over the entire planning horizon. At each stage, system components may be exposed to or damaged by wildfire, necessitating corrective operational responses that simultaneously prioritize the service of higher-value loads and minimize overall operational costs. The framework further incorporates corrective adjustments to account for the progressive effects of wildfire propagation in subsequent stages. In this manner, the optimization systematically evaluates the full set of preventive and corrective strategies across all possible trajectories, ensuring coordinated and adaptive decision-making throughout the horizon.

Equation (1) is designed to minimize the expected value of wildfire risk $R_{fire,n}$, maximize weighted load delivery $D_{Tot,n}$, and account for generation operational costs $C_{g,n}(p_{g,n})$ for every node $n \in N$. The parameter $\alpha \in [0\ 1]$, represents the trade-off between prioritizing load delivery and mitigating wildfire risk. The expected value formulation systematically accounts for all potential wildfire impacts on power system components across various spread scenarios, incorporating their associated probabilities at each stage and for all nodes within the multistage optimization process.

$$Min\ \mathbb{E}[\sum_{n\in N}(\alpha R_{fire,n} - (1-\alpha)(D_{Tot,n} - \sum_{g\in G} C_{g,n}(p_{g,n}))] \tag{1}$$

The extended nodal formulation of (1) is represented by (2). where $\rho_n$ represents the probability associated with the occurrence of each spread scenario at a given node in the decision tree.

$$Min\ [\sum_{n\in N}(\rho_n(\alpha R_{fire,n} - (1-\alpha)(D_{Tot,n} - \sum_{g\in G} C_{g,n}(p_{g,n})))] \tag{2}$$

In terms of optimization variables, the decision of whether to de-energize generators, lines, and buses is represented by binary decision variables $u_{*,n} \in \{0,1\}$, where 1 indicates that the component is energized and 0 indicates it is not at each node $n$. The decision variables $u_{g,n}, u_{e,n}, u_{i,n}$ correspond to the status of generators, lines, and buses, respectively.

Load shedding is modeled as a continuous decision variable $l_{d,n} \in [0\ \ 1]$, representing the fraction of the load that is de-energized.

In this approach, the risk of triggering a wildfire by electrical components at each node $n$ is modeled as described in (3) and is represented by $R_{fire,n}$. This node-specific risk definition accounts for the assumption that wildfire risk varies across stages and components, depending on power flow and environmental conditions in each scenario.

The total amount of weighted load delivered at node $n$, $D_{Tot,n}$, is expressed by (4), where $D_{d,n}$ is the amount of load served under normal operating conditions at node $n$, and $\omega_{d,n}$ is a weight used to prioritize certain critical loads during a wildfire event.

The energization of generators, loads, and lines depends on the energization status of the buses they are linked to. This requirement is enforced through the constraints outlined in (5), (6), and (7).

$$R_{fire,n} = \sum_{d\in D} l_{d,n} R_{d,n} + \sum_{g\in G} u_{g,n} R_{g,n} + \sum_{e\in E} u_{e,n} R_{e,n} + \sum_{i\in B} u_{i,n} R_{i,n} \tag{3}$$

$$D_{Tot,n} = \sum_{d\in D} l_{d,n}\omega_{d,n} D_{d,n} \tag{4}$$

$$u_{i,n} \geq l_{d,n} \quad \forall d \in B_i^D\ \forall i \in B \tag{5}$$

$$u_{i,n} \geq u_{g,n} \quad \forall g \in B_i^G\ \forall i \in B \tag{6}$$

$$u_{i,n} \geq u_{e,n} \quad \forall e \in B_i^e\ \forall i \in B \tag{7}$$

Equations (8) to (10) represent the DC power flow equations for branches. The characteristics of the line are defined by its thermal limit, $T_e$, and the susceptance, $b_{e,n}$, for node $n$ in the decision tree.

In (8) to (10), the term $u_{e,n}$ denotes binary decision variables that capture the effects of the PSPS decision on the line flows. PSPS variables determine the operational status of components, thereby influence the power flow calculations. Constraint (10) indicates that power flow is restricted to remain within the thermal power flow limit, and $u_{e,n}$ is incorporated to ensure that the branch power flow is zero if the line is de-energized.

$$P_{e,n} \leq -b_{e,n}(\theta_{i,n} - \theta_{j,n} + \theta^{max}(1-u_{e,n})) \quad \forall i \in B, \forall j \in B, e \in E(i,j) \tag{8}$$

$$P_{e,n} \geq -b_{e,n}(\theta_{i,n} - \theta_{j,n} + \theta^{min}(1-u_{e,n})) \quad \forall i \in B, \forall j \in B, e \in E(i,j) \tag{9}$$

$$-T_e u_{e,n} \le P_{e,n} \le T_e u_{e,n} \qquad (10)$$

Maintaining the stability of an electrical system necessitates balancing the generated power with the electricity demand at each bus within the network. Constraint (11) in the model represents power balance equations. For every bus $i$ in the network, there exists an equation for the balance of real power, where the total injected supply must equal the withdrawals.

In (11), the term $l_{d,n}$ represents the impact of PSPS decisions on the load in power balance equations of each node.

$$\sum\nolimits_{g \in B_i^G} p_{g,n} = \sum\nolimits_{e \in B_i^e} P_{e,n} + \sum\nolimits_{d \in B_i^d} l_{d,n} D_{d,n} \qquad (11)$$
$$\forall i \in B, e \in E(i,j)$$

Equation (12) depicts the active power generation of generator $g$ in node $n$ with respect to the one in the preceding stage of decision tree. This constraint facilitates adjustments in the generation of each generator by adding only a generation difference $\Delta p_{g,n}$ to the generation level from the previous stage $p_{g,a(n)}$, where $a(n)$ represents the parent node of $n$.

$$p_{g,n} = p_{g,a(n)} + \Delta p_{g,n} \qquad (12)$$
$$\forall g \in G$$

The variables in (12) are considered state variables for power generation, playing a crucial role in the multistage optimization process. Equation (13) defines the generator ramping rate, ensuring that generators do not exceed the maximum ramp-up or ramp-down rate. This constraint limits the rate at which generators can change their output levels during the transition from one stage to another one.

$$\Delta p_g^{min} \le \Delta p_{g,n} \le \Delta p_g^{max} \qquad (13)$$
$$\forall g \in G$$

Constraint (14) defines the lower and upper bounds for real power production of generators in each node, where $u_{g,n}$ is regarded as a decision variable, determined during PSPS, as it may be necessary to shut down certain generators.

$$u_{g,n} P_g^{min} \le p_{g,n} \le u_{g,n} P_g^{max} \qquad (14)$$
$$\forall g \in G$$

## III. Wildfire-Driven Decision Tree Construction

In this section, an algorithm is proposed to automatically generate a decision tree based on wildfire spread scenarios, identifying electrical components exposed to the wildfire. An essential aspect of multistage optimization is the capability to make decisions at each stage of the process, while accounting for the entire problem horizon represented as a decision tree. Decision trees for wildfire events in power systems should be constructed based on the progression of the wildfires. To ensure robust decision-making, all possible scenarios must be considered based on the wildfire's spread directions and the power system's geographical position. Accurately capturing these dynamics is vital for developing effective response strategies and optimizing system resilience. In this context, Algorithm 1 is introduced to accurately capture and automatically construct the decision tree, enabling informed decision-making. When a wildfire occurs, the algorithm determines the probability of failure of various electrical components (e.g., power lines) exposed to the wildfire, based on the geographic location of the wildfire in relation to the locations of components.

Considering factors such as distance, weather conditions, fuel data (e.g., land type), and historical wildfire data, probabilities can be assigned to each scenario.

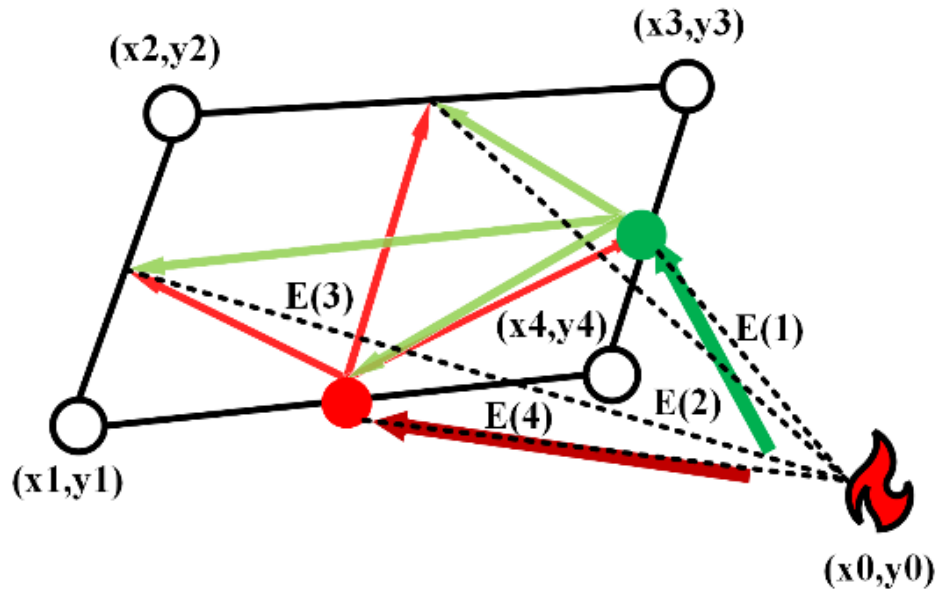

**Figure 2** Wildfire propagation scenarios example based on algorithm 1

Figure 2 presents a dynamic assessment of wildfire impacts on a four-bus electrical system over two stages, based on the results generated by Algorithm 1. The wildfire's initial location is denoted by coordinates $(x_0, y_0)$ while the locations of the four buses are specified by their respective coordinates $(x_1, y_1)$ to $(x_4, y_4)$. If power lines intersect, the intersection point is also considered as $(x_i, y_i)$ point, and each segment of the intersecting lines is treated as a separate line for the algorithm's analysis. Based on the proximity of wildfire to the transmission lines connecting these points (i.e. buses), the algorithm aims to identify the lines that will be the first to be exposed to wildfire. Each exposed line is then treated as a new starting point, as illustrated in Figure 2 by their midpoints with red and green dots. Once the new impact points are identified, the wildfire location is updated, allowing the algorithm to be rerun with the revised impact points to determine new impact component scenarios. This recursive process builds a scenario tree that simulates the wildfire's progression.

**Algorithm 1: Automated Decision Tree Algorithm for Wildfire-Exposed Electrical Components**

**Require:** Equation of lines = $\{Y = m_1 X + d_1 \;\; x_{s1} \le X \le x_{e1};\; Y = m_2 X + d_2 \;\; x_{s2} \le X \le x_{e2}; \ldots;\; Y = m_k X + d_k \;\; x_{sk} \le X \le x_{ek}$

1: **Input:** fire point = $(x_0, y_0)$
2: **for** $i = 1$ *to* $k$ **do**
3: $S(i) = 0$
4: $x_{middle} = \frac{x_{i1} + x_{i2}}{2}$
5: $y_{middle} = \frac{m_i x_{si} + d_i + m_i x_{ei} + d_i}{2}$
6: $E(i) = \frac{y_0 - y_{middle}}{x_0 - x_{middle}}(X - x_0) + y_0 \quad x_0 \le X \le x_{middle}$
7: **for** $j$: 1 *to* $k$ **do**
8: **Evaluate: X from** E(i) = $m_j X + d_j$
9: **if** $x_0 \le x \le x_{middle}$ **And** $x_{sj} \le x \le x_{ej}$ **then**
10: $S(i) = S(i) + 1$
11: **end if**
12: **end for**
13: **if** $S(i) = 1$ **then**
14: **Consider** i
15: **end if**
16: **end for**

In Algorithm 1, the first loop begins by calculating the midpoint $(x_{middle}, y_{middle})$ of each line based on the line equation, derived from the power system configuration. Subsequently, using the wildfire location, the algorithm determines the line equation $E(i)$ connecting the wildfire location to the calculated midpoints. Within the inner loop, the algorithm evaluates the number of hit points between the wildfire and all midpoints, adhering to the constraints defined by the power system's topology. If the number of hit points is equal to one, the scenario is considered a viable possibility; otherwise, it is disregarded. This allows the algorithm to identify the first power lines to be impacted by wildfire at each stage. This process is repeated iteratively based on the required number of stages for the optimization.

## IV. Solution Algorithm

The multistage stochastic problem formulated in this paper is considered a large-scale optimization problem. In particular, as the number of stages in the multistage optimization increases, the number of variables and constraints grows significantly. Additionally, due to the interdependence of decision variables across stages and the high number of mixed-integer variables involved, the problem becomes highly complex. As a result, conventional solvers are often inadequate for efficiently solving the problem. The solution proposed for this problem offers significant advantages for solving mixed-integer multistage optimization problems. By employing the Stochastic Dual Dynamic Programming (SDDP) approach, along with advanced techniques such as generalized conjugacy cuts and regularization, the algorithm is capable of efficiently managing complex optimization problems. These unique approaches enable the global optimization of a broad class of multistage stochastic problems, providing a robust framework for addressing challenges inherent in large-scale, and mixed-integer models.

The proposed framework inherits the scalability properties of SDDP-based methods by exploiting stagewise decomposition and recombining scenario tree structure. Consequently, the computational burden grows primarily with the number of stages, rather than exponentially with the number of scenarios. Moreover, theoretical results in the SDDP literature indicate that the number of iterations required to obtain an optimal first-stage solution scales approximately linearly with the number of stages $T$ [24]. The proposed algorithm preserves this favorable scaling behavior while extending SDDP to mixed-integer multistage settings through regularization and cut generation. This makes the framework well suited for large-scale applications with long planning horizons, such as preventive–corrective optimization of power systems under wildfire risk.

In this framework, first, all state variables and internal variables vectors based on the optimization formulation in Section II are defined. Subsequently, the optimization problem is reformulated using the SDDP approach [23]. Finally, the proposed algorithm for solving the optimization problem is presented. Within the optimization framework, the vector $v_n$ is defined, which includes all state variables relevant to the system's operation. This vector comprises state variables including the power generation of each generator $(p_{g,n})$, as well as generator status $(u_{g,n})$, busbar status $(u_{i,n})$, load shedding $(l_{d,n})$, and transmission line status $(u_{e,n})$, all of which are influenced by PSPS decisions.

All other variables, such as voltage phase angle and lines power flows, defined in the optimization problem and excluding state variables, are considered internal variables. These internal variables are collectively represented by the vector $w_n$ within the optimization framework.

In this study, to systematically formulate the optimization problem, the objective function is denoted as $f_n$ in (15), which is originally defined by (2). Consequently, in the solution algorithm, we will refer to$f$ without explicitly referencing (2). Additionally, to utilize the solution algorithm effectively, $v_{a(n)}$ that is a parent vector of $v_n$ in the decision tree is redefined within the objective function rather than as part of the constraints. In particular, this redefinition entails substituting $p_{g,a(n)}$ from Constraint (12) into the objective function.

$$\min_{(v_n,w_n)\in F_n}[\sum_{n\in N}\rho_n f_n(v_{a(n)},w_n,v_n)] \tag{15}$$

The recursive formulation of (15) is represented by (16):

$$\begin{aligned}Q_n(v_{a(n)},w_n,v_n) = \min_{(v_n,w_n)\in F_n}\ &[f_n(v_{a(n)},w_n,v_n)\\ &+\sum_{m\in C(n)}\rho_{nm}Q_m(v_n)]\end{aligned} \tag{16}$$

Where $n \in N$ represents a non-leaf node and $Q_n$ denotes the value function of node *n*. The term on the right-hand side of (16) is known as the nodal problem of node *n*. Its objective function encompasses the nodal cost function and the expected cost-to-go function. Where the expected cost-to-go function is defined in (17). The expected cost-to-go function becomes zero at a leaf node, since there are no child nodes $(C(n) = 0)$. In this context, $\rho_{nm}$ represents the transition probability that node *m* is realized, given that its parent node *n* has been realized.

$$\mathcal{Q}_n(v_n) = \sum_{m\in C(n)}\rho_{nm}Q_m(v_n) \tag{17}$$

For the proposed solution algorithm, it is necessary to define three subproblem oracles. These oracles are essential for describing the algorithms and conducting complexity analysis. A subproblem oracle is an entity that, given subproblem information and current algorithm information, generates a solution to the subproblem.

**Definition 1:** Forward Step Subproblem Oracle for Non-Root Nodes $O_n^f$ is defined in (18).

$$\begin{aligned}\min_{(v,w)\in F_n, z\in V_{a(n)}}\ &[f_n(z,w,v)+\sigma_n\psi_n(v_{a(n)}-z)\\ &+\Theta_n(v)]\end{aligned} \tag{18}$$

$O_n^f$ takes $(v_{a(n)}, \Theta_n(v))$ as input, and outputs an optimal solution $(v_n, w_n, z_n)$ for $n \neq r$.

**Definition 2:** Backward Step Subproblem Oracles for Non-Root Nodes $O_n^B$ is presented in (19).

$$\begin{aligned}\max_{(\lambda,\rho)\in U_n}\ \min_{(v,w)\in F_n, z\in V_{a(n)}}\ &[f_n(z,w,v)+\langle\lambda, v_{a(n)}-z\rangle\\ &+\sigma_n\psi_n(v_{a(n)}-z)+\Theta_n(v)]\end{aligned} \tag{19}$$

Similarly, $O_n^B$ takes $(v_{a(n)}, \Theta_n(v))$ as input, and outputs an optimal solution $(v_n, w_n, z_n; \lambda_n, \rho_n)$ for $n \neq r$.

**Definition 3:** Subproblem Oracle for the Root Node $O_r$, is provided in (20).

$$\min_{(v,w)\in F_n,\ z\in V_{a(n)}}[f_r(v_{a(r)},w,v)+\Theta_r(v)] \tag{20}$$

Here, $O_r$ takes $\Theta_r(v)$ as input, and outputs an optimal solution $(v_r, w_r)$.

Assume $(v_n^i, w_n^i)_{n\in N}$ are feasible solutions to the regularized nodal problem in *i-th* iteration. Then, the under-approximation of the expected cost-to-go function is defined recursively from the leaf nodes to the root node, and inductively for $i \in N$ as shown in (21) to (24):

$$\underline{Q}_n^i(v) = \max\{\underline{Q}_n^{i-1}(v), \sum_{m\in C(n)} \rho_{nm} C_m^i(v|\hat{\lambda}_m^i, \hat{\rho}_m^i, \underline{q}_m^i)\} \quad \forall\, v \in V_n \tag{21}$$

$$C_m^i\left(v\middle|\hat{\lambda}_m^i, \hat{\rho}_m^i, \underline{q}_m^i\right) = -\langle\hat{\lambda}_m^i, v_n^i - v\rangle - \hat{\rho}_m^i \psi_m(v_n^i - v) + \underline{q}_m^i \tag{22}$$

$$\left(\hat{v}_m^i, \widehat{w}_m^i, \hat{z}_m^i; \hat{\lambda}_m^i, \hat{\rho}_m^i\right) = O_m^B\,(v_n^i, \underline{Q}_m^i) \tag{23}$$

$$\underline{q}_m^i = f_m(\hat{z}_m^i, \widehat{w}_m^i, \hat{v}_m^i) + \langle\hat{\lambda}_m^i, v_n^i - \hat{z}_m^i\rangle + \hat{\rho}_m^i \psi_m(v_n^i - \hat{z}_m^i) + \underline{Q}_m^i(\hat{v}_m^i) \tag{24}$$

(25) and (26) illustrate the over-approximation of the regularized expected cost-to-go functions, which are employed in the sampling and termination processes of the proposed nested decomposition and dual dynamic programming algorithms for $i \in N$ at the root node *r*. This over-approximation of the regularized expected cost-to-go function is defined recursively, from leaf nodes to the child nodes of the root node, and inductively for $i \in N$.

$$\overline{Q}_n^i(v) = \min\{\overline{Q}_n^{i-1}(v), \rho_{nm}(\overline{q}_m^i + \sigma_m||v - v_n^i||)\} \tag{25}$$

$$\overline{q}_m^i = f_m\left(z_m^i, w_m^i, v_m^i + \sigma_m \psi_m(v_n^i - \hat{z}_m^i) + \overline{Q}_m^i(v_m^i)\right) \tag{26}$$

The algorithm used to solve the optimization problem is detailed in Algorithm 2. This algorithm comprises three main steps: a forward step, a backward step, and a root node update step. In the forward step, for a node at a given stage, the algorithm identifies a child node in the next stage that exhibits one of the largest approximation gaps. The state variable of this selected child node is then used as the state variable for the current stage in the iteration. Due to the stagewise independence of the problem, the backward step at each stage only generates necessary cuts for nodes within the recombining tree structure.

**Lemma 1:** The nodal cost function $f_n$ is nonnegative, and the feasibility set $F_n$ is compact.

**Proof:** The nodal cost function $f_n$ includes two components: risk and operational cost. The risk component is always nonnegative, as it is either positive or zero by definition. Similarly, the operational cost is inherently positive due to the nature of operational expenses in power systems. Together, these ensure that $f_n$ is nonnegative.

To establish the compactness of the feasible set $F_n$, we rely on the structure of the optimization problem. Compactness requires $F_n$ to be both bounded and closed. The constraints of the optimization problem ensure that all feasible solutions lie within a bounded region of the decision space, as physical and operational limits of the power system impose finite upper and lower bounds.

**Algorithm 2: Multistage Solution Algorithm**

**Require:** Scenario tree from section III and subproblem oracles $O_n^f, O_n^B, and\ O_r\ n \neq r, and\ \varepsilon > 0$

1: **Initialize** $i \leftarrow 1$; $\underline{Q}_t^0 \leftarrow 0, \forall t, \overline{Q}_t^0 \leftarrow +\infty, \forall t \leq T-1$; $\overline{Q}_T^0 \leftarrow 0$
2: **Evaluate** $(v_0^1, w_0^1) = O_r(0)$
3: **Set** LOWERBOUND$\leftarrow f_r(v_{a(r)}, w_0^1, v_0^1)$, UPPERBOUND$\leftarrow +\infty$
4: **while** UPPERBOUND - LOWERBOUND$> \varepsilon$
5: **for** *t=1* to T-1 **do**
6: **for** $n \in \widetilde{N}(t)$ **do**
7: **Evaluate** $(v_n^i, w_n^i, z_n^i) = O_n^f(v_{t-1}^i, \underline{Q}_t^{i-1})$
8: **Calculate the gap** $\gamma_n^i = \overline{Q}_t^{i-1}(v_n^i) - \underline{Q}_t^{i-1}(v_n^i)$
9: **end for**
10: **Select any** $n^*(t) \in \{n \in N(t): \gamma_n^i \geq \gamma_{n'}^i, \forall n' \in \widetilde{N}(t)\}$**, and let** $v_t^i \leftarrow v_{n^*(t)}^i$
11: **end for**
12: **for** t=T to 1 **do**
13: **Update** $\underline{Q}_t^i$ and $\overline{Q}_t^i$ **using** (21) and (25)
14: **for** $n \in \widetilde{N}(t)$ **do**
15: **Evaluate** $(\hat{v}_n^i, \widehat{w}_n^i, \hat{z}_n^i; \hat{\lambda}_n^i, \hat{\rho}_n^i) = O_n^B(v_{t-1}^i, \underline{Q}_t^i)$
16: **Calculate** $C_n^i, \underline{q}_n^i, \overline{q}_n^i$ **using** (22), (24), (26)
17: **end for**
18: **end for**
19: **Update** $\underline{Q}_0^i, \overline{Q}_0^i$ **using** (21) and (25)
20: **Evaluate** $(v_0^{i+1}, w_0^{i+1}) = O_r(\underline{Q}_0^i)$
21: **Update** LOWERBOUND$\leftarrow f_r(v_{a(r)}, w_0^{i+1}, v_0^{i+1}) + \underline{Q}_0^i(v_0^{i+1})$
22: **if** UPPERBOUND$>f_r(v_{a(r)}, w_0^{i+1}, v_0^{i+1}) + \overline{Q}_0^i(v_0^{i+1})$ **then**
23: **Update** UPPERBOUND$\leftarrow f_r(v_{a(r)}, w_0^{i+1}, v_0^{i+1}) + \overline{Q}_0^i(v_0^{i+1})$
24: **Set** $(v_0^*, w_0^*) = (v_0^{i+1}, w_0^{i+1})$
25: **End if**
26: $i \leftarrow i + 1$
27: **End while**

**Lemma 2:** The nodal cost function $f_n$ is lower semicontinuous (l.s.c.).

**Proof:** The objective function $f_n$ consists of three components: $R_{fire,n}, D_{Tot,n}$, and $C_{g,n}(p_{g,n})$. To prove that $f_n$ is lower semicontinuous, the continuity of each component is examined. Both $D_{Tot,n}$ and $C_{g,n}(p_{g,n})$ are linear functions, which are continuous by definition. A continuous function is always lower semicontinuous, so both terms satisfy the l.s.c. condition.

The risk term $R_{fire,n}$ defined in (3) consists of four components. First, $l_{d,n}R_{d,n}$ that this term is continuous, as $l_{d,n}$ is continuous functions and $R_{d,n}$ is constant in each node *n*. Therefore, it is l.s.c. $u_{g,n}R_{g,n}$, $u_{e,n}R_{e,n}$, and $u_{i,n}R_{i,n}$ terms are Lipschitz continuous, satisfying the condition (27).

$$\left|R_{fire,n(z,w,v)} - R_{fire,n(z'w,v)}\right| \leq L\,|z - z'| \tag{27}$$

Where $L$ is the Lipschitz constant. Lipschitz continuous functions are a subset of continuous functions and are therefore also lower semicontinuous. Since all individual components of $R_{fire,n}$ are lower semicontinuous, $R_{fire,n}$ is also lower semicontinuous.

**Theorem:** If, for every node $n \in N$ in the feasibility set, $F_n$ is compact, the nodal cost function $f_n$ is nonnegative and lower semicontinuous (l.s.c.), and the sum $\sum_{n \in N} \rho_n f_n$ is a proper function—i.e., there exists $(v_n, w_n) \in F_n$ for all nodes $n \in N$ such that $\sum_{n \in N} \rho_n f_n \left(v_{a(n)}, w_n, v_n\right) \leq +\infty$, then the proposed algorithmic framework guarantees the existence of an optimal solution to the optimization problem.

**Proof:** The proof can be found in [23].

## V. Simulation and Result

The proposed automated decision-support methodology is validated by applying it to the IEEE 30-bus system. The power generation data are presented in TABLE I.

**TABLE I**

Generator Data

| Unit | Power (MW) | | Cost ($/MW) |
|---|---|---|---|
| | Min | Max | |
| $G_1$ | 50 | 200 | 20 |
| $G_2$ | 20 | 80 | 17.5 |
| $G_3$ | 15 | 50 | 10 |
| $G_4$ | 10 | 35 | 32.5 |
| $G_5$ | 10 | 30 | 30 |
| $G_6$ | 12 | 40 | 30 |

As depicted in Figure 3, it is assumed that wildfire originates in the center of the system, with all potential spread scenarios generated by Algorithm 1. A region within the system is identified as a fire-prone area. The high-risk area, along with the associated system components it contains, is depicted in Figure 3. In this simulation, it is assumed that only power lines are impacted by wildfires. The wildfire spread scenarios at the initial stage, represented by large red arrows, illustrating the potentially affected power lines. In the subsequent stage, the probability of wildfire spread scenarios is calculated based on the initial realized scenario, with potential future scenarios represented by thinner red arrows. This process is repeated till the last stage of wildfire scenarios. The probability of each scenario occurring is influenced by several factors, including proximity to the fire, vegetation density, and wind speed and direction. Considering these factors, the scenario tree depicted in Figure 4 is generated, with each scenario assigned a corresponding probability of occurrence.

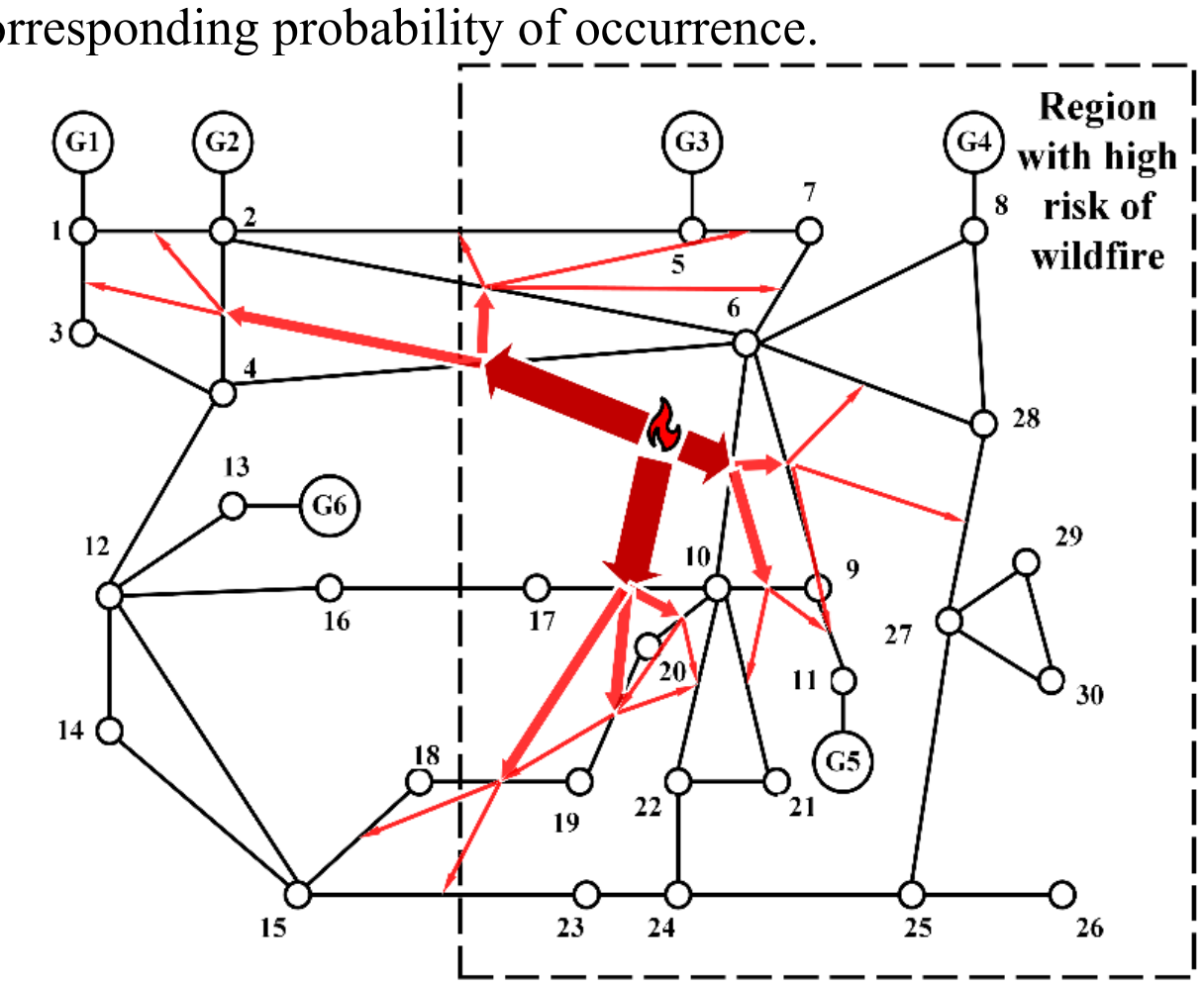

**Figure 3.** IEEE 30-bus system with wildfire propagation scenarios and high-risk wildfire regions.

It is assumed that regions and components outside the high-risk area have zero risk of wildfire ignition. Conversely, transmission lines entering the high-risk wildfire region are associated with high levels of risk, reflecting their greater potential to contribute to wildfire incidents. Without loss of generality, generators, buses, and loads within the high-risk wildfire region are considered to have a low probability of initiating an ignition. Additionally, a ramping rate of 20 MW/hour is considered for all generators, aligned with the hourly intervals of wildfire progression.

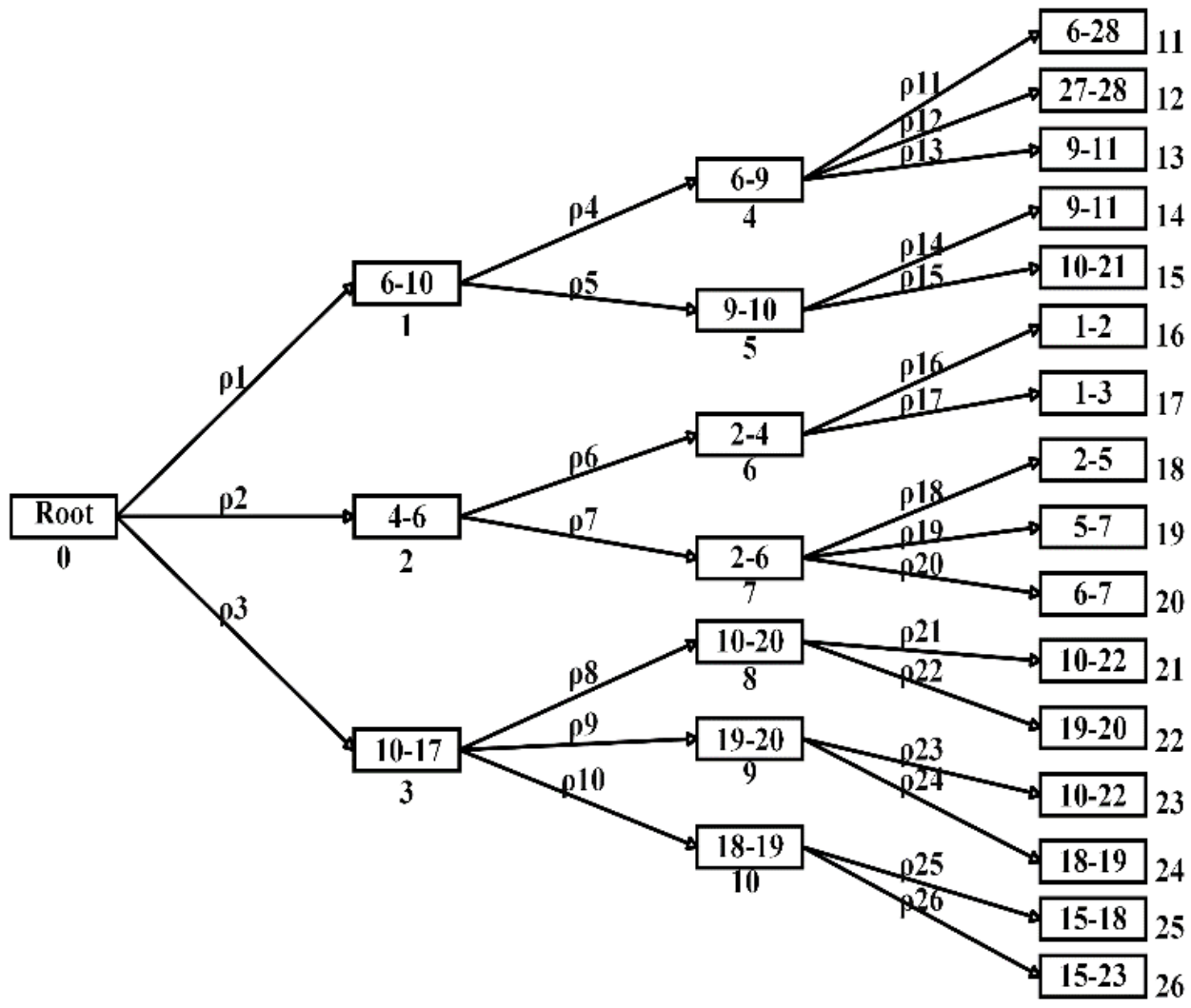

**Figure 4**. Multistage decision tree for the IEEE 30-bus system based on wildfire locations

The optimization was carried out using two distinct decision-making approaches: a multistage optimization (MO) framework and a series of single-stage optimization (SSSO) framework. The MO framework incorporates future contingency scenarios, enabling coordinated preventive and corrective decision-making, automating sequential operations, and enhancing the effectiveness of resilience planning. In contrast, the SSSO approach determines preventive actions based solely on the system's initial state, while corrective actions are made independently for each scenario after a component failure, without considering the possibility of future failures. In both cases, the same operational constraints, including ramping rates, are applied. Ramping rates limit the amount of change in power generation between stages compared to the previous stage.

The risk of wildfires caused by power lines is influenced by the power flow through transmission lines. Therefore, any actions—whether preventive or corrective—that affect power flow will, in turn, influence wildfire risk over time. In practice, while we may be able to reduce the risk of endogenous wildfires (those initiated by the power system), we have no control over exogenous wildfires, which can ignite spontaneously at any location within fire-prone regions. As these exogenous wildfires progress, they can disrupt system topology and alter power flows—thereby impacting the wildfire risk posed by power lines throughout the network. Consequently, preventive actions should be designed to preserve flexibility for responding to future risks rather than constraining the system's adaptability. In this context, the MO framework evaluates wildfire risk across all decision tree nodes over the entire optimization horizon, allowing it to reflect dynamic and uncertain real-world conditions. In contrast, the SSSO approach bases preventive decisions solely on the current state of the system, without accounting for future contingency scenarios or evolving wildfire risks.

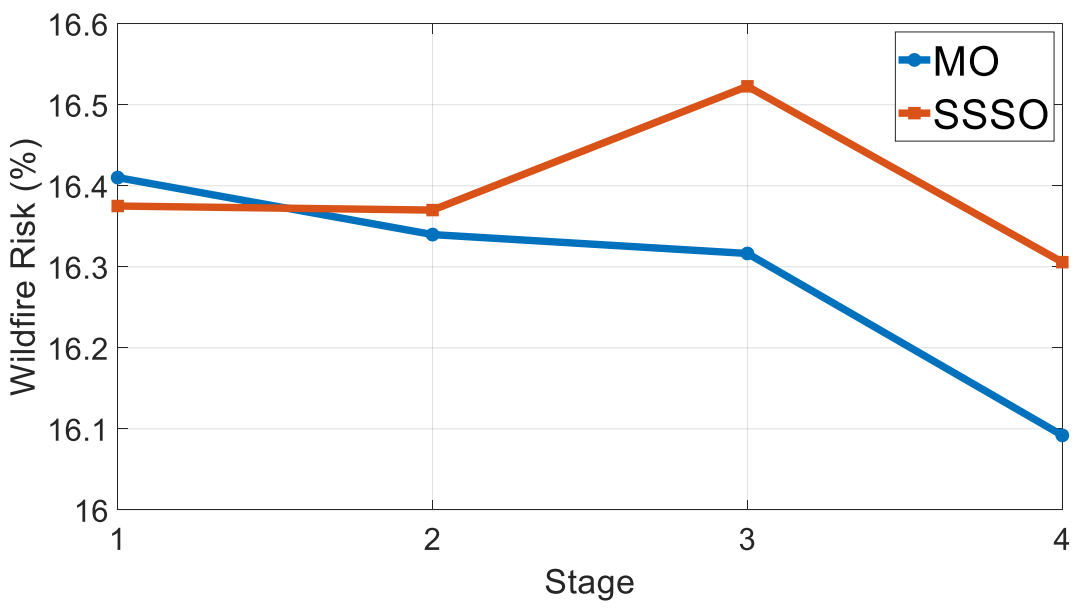

**Figure 5** Expected risks at each stage in MO and SSSO following preventive decisions.

The MO and SSSO frameworks are applied to identical scenarios to enable a direct comparison of their performance. In the first case study, it is assumed that lines 6–8 and 8–28 pose approximately equal wildfire ignition risks. The SSSO approach, relying solely on the system's initial conditions, de-energizes line 8–28 without considering potential future contingencies—limiting its adaptability. In contrast, the MO framework evaluates the long-term implications of early-stage decisions. It identifies that de-energizing line 8–28 in the first stage could increase risk in later stages. To mitigate this, the MO approach strategically avoids disconnecting line 8–28 and instead de-energizes line 6–8, thereby minimizing cumulative wildfire risk over the entire planning horizon.

Figure 5 illustrates the wildfire risk associated with each stage under both the SSSO and MO frameworks, based on the preventive decisions made by each approach. The results show that while the SSSO approach yields a marginally lower risk in the first stage, it does not result in a lower expected overall risk due to its initial preventive decision. In contrast, the MO approach, although showing a slightly higher risk in the initial stage, accounts for wildfire risks in future stages. As a result, the MO approach leads to a more comprehensive reduction in risk across the entire time horizon.

In Figure 6, a comparison of the operational cost of final stage nodes for both MO and SSSO is presented. As shown, in most scenarios, the associated costs in the final stages of optimization in MO compared to SSSO are lower or approximately equal.

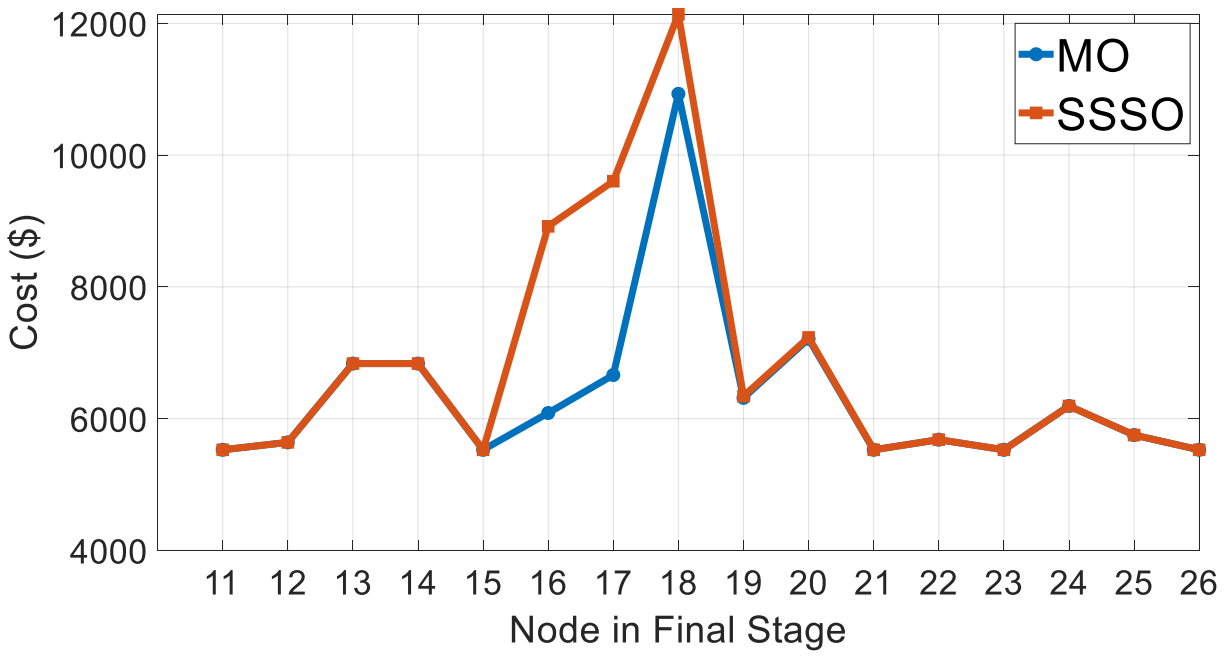

**Figure 6.** Associated costs of the final stage nodes for both MO and SSSO.

To illustrate the costs and power profiles of generators across all stages for specific scenario paths, we analyzed paths 0-2-6-16 and 0-2-6-17 in the decision tree. Figure 7 illustrates the costs associated with all stages in the decision tree for paths 0-2-6-16 and 0-2-6-17.

For the path 0-2-6-16, Figure 7(a) shows that the operational costs in the second and third stages for MO are slightly higher than those for the SSSO. In specific, the difference is about 2% in the second stage and 4.2% in the third stage. However, there is a substantial difference in the associated cost for the last stage, being approximately 47% less for MO compared to SSSO. This shows that while the initial-stage decisions in the MO approach may appear more conservative, the performance significantly improves as the wildfire evolves, ultimately outperforming the SSSO strategy in later stages.

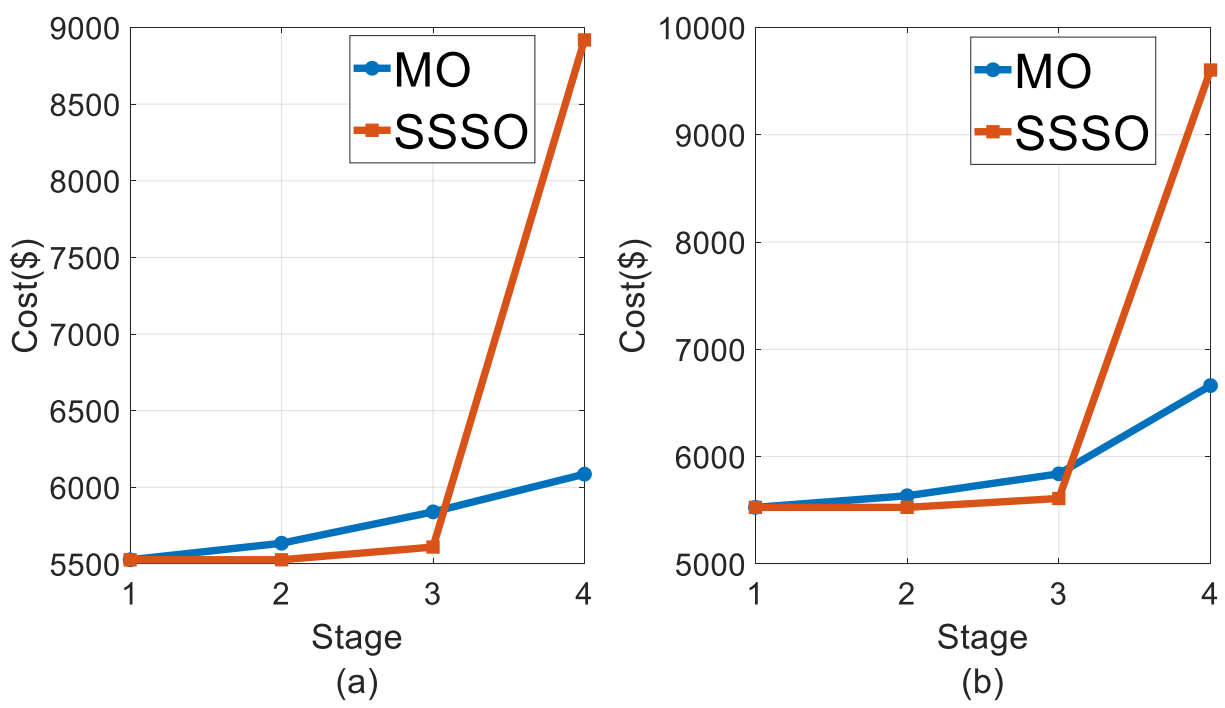

**Figure 7.** Associated nodal costs for all stages for (a) Path 0-2-6-16 and (b) Path 0-2-6-17.

Similarly, for the path 0-2-6-17 shown in Figure 7(b), a difference of approximately 43% is observed at the final stage. This highlights that the situation could be much worse in the last stage if future scenarios are not considered in decision making process. As shown in Figure 6, a similar trend is observed for nodes 18, 19, and 20.

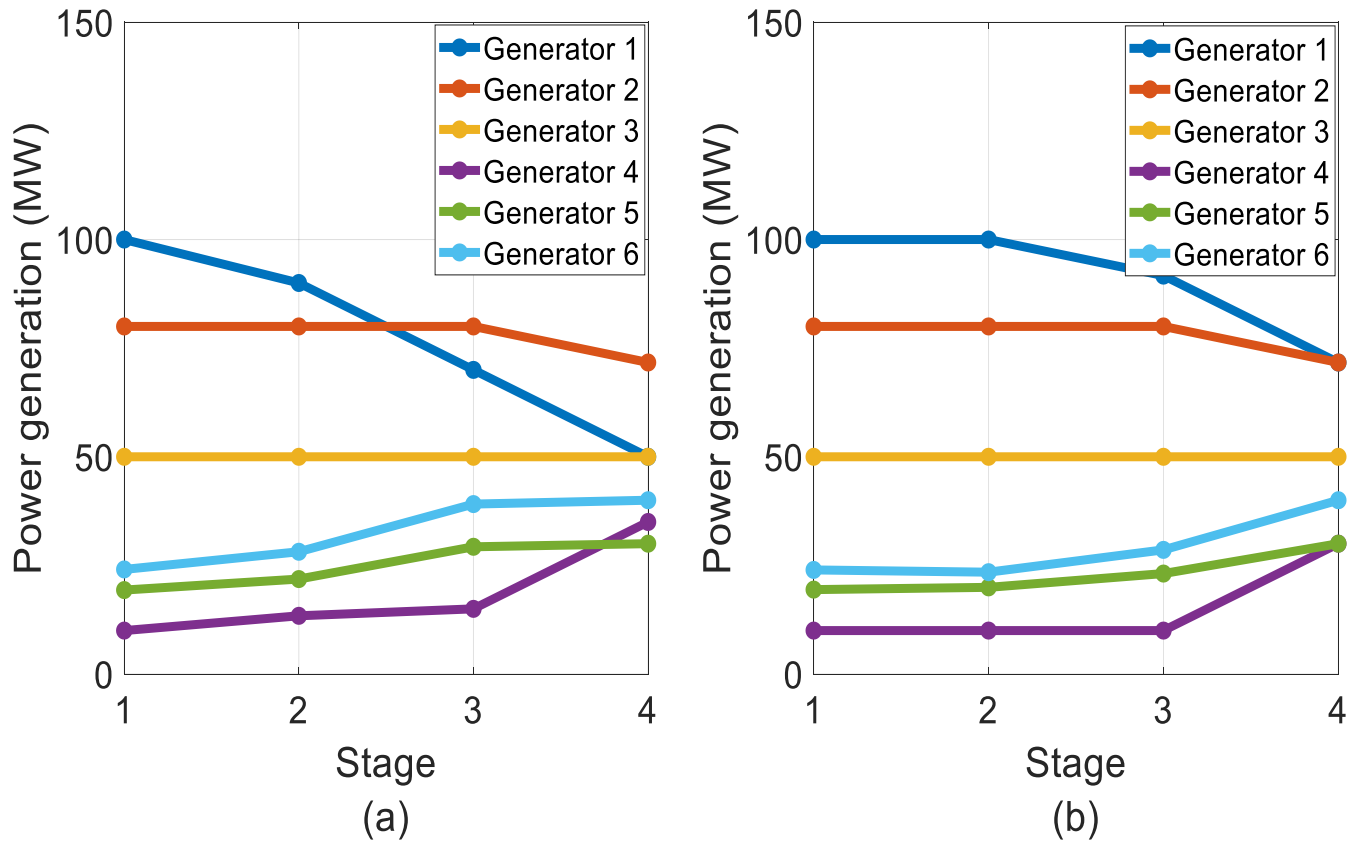

**Figure 8.** Power generation for all generators across stages for Path 0-2-6-17 for (a) MO and (b) SSSO.

Figure 8 illustrates the generators' output power for both MO (7a) and SSSO (7b) frameworks along the path 0-2-6-17. In this scenario, Generator 1 loses one of its connections, requiring a reduction in power output to maintain thermal limits on the remaining power lines. The multistage optimization framework anticipates this event and adjusts power allocation in earlier stages, enabling a controlled reduction in generation. As a result, during the final stage, the generator operates at its minimum allowable limit, leading to significantly lower costs compared to the SSSO approach. Although the initial operational costs in the multistage optimization framework may be higher than those in the SSSO, the multistage approach offers significant advantages as the wildfire progresses and affects power lines. Specifically, it results in lower overall operational costs and a considerable reduction in wildfire risk compared to the conventional SSSO framework. This underscores the long-term advantages of automated, anticipatory, and adaptive decision-making in managing dynamic and uncertain conditions of power systems and wildfire risks.

To demonstrate the scalability of the proposed optimization model and solution algorithm, the framework is applied to the IEEE 300-bus test system, which provides a sufficiently large number of busbars to support multistage formulations with increasing temporal depth and scenario resolution. A wildfire event is assumed to occur at a selected interior location of the network to reflect realistic large-scale operating conditions. The optimization problem is solved under multiple configurations of the multistage decision structure, in which both the number of stages and the number of scenario nodes are progressively increased. All simulations were performed on a workstation equipped with a 12th-generation Intel® Core™ i7-12700 CPU and 32 GB of RAM. Table II reports the optimal objective values and computational times corresponding to different combinations of stages and scenario nodes. The results demonstrate that the proposed solution algorithm remains computationally efficient as the multistage problem complexity grows, confirming its scalability for large-scale power system applications.

**TABLE II**

Objective values and solution times for the IEEE 300-bus system under different multistage configurations.

| Number of Stages Considered | Number of Decision Tree Nodes | Optimal Value | Computational Time (s) |
|---|---|---|---|
| 2 | **24** | 156522 | **291** |
| 3 | **70** | 241521 | **842** |
| 4 | **150** | 342154 | **1867** |

## VI. Conclusion

In conclusion, this paper presents a preventive-corrective multistage optimization framework designed to enhance the resilience of power systems during wildfire events. The key contributions of this study include the development of a multistage optimization model that considers PSPS actions prior to wildfire occurrence, predicts the future states of power systems during wildfire progress, and implements corrective actions to address potential system failures in an automated and reliable manner. Additionally, a novel wildfire-driven algorithm for automatically generating decision trees based on the geographical positions of wildfires and the power network is introduced. An SDDP algorithm is utilized to solve the optimization problem and achieve a global optimum with improved computational efficiency and scalability. The results demonstrate the effectiveness of the proposed automated methodology compared to single-stage operational solutions, effectively minimizing operational costs, wildfire risk, and load outages within the optimization horizon.